\documentclass[aps,prb,reprint,showpacs,superscriptaddress]{revtex4-1}

\usepackage[dvips]{graphicx}
\usepackage{amsmath}

\begin{document}

\title{Topographic and electronic contrast of the graphene moir\'{e} on Ir(111) probed by scanning tunneling microscopy and non-contact atomic force microscopy}

\author{Zhixiang Sun}
\affiliation{Condensed Matter and Interfaces, Debye Institute for Nanomaterials Science,
Utrecht University, PO Box 80000, 3508 TA Utrecht, the Netherlands}
\author{Sampsa K. H\"{a}m\"{a}l\"{a}inen }
\author{Jani Sainio}
\author{Jouko Lahtinen}
\affiliation{Department of Applied Physics, Aalto University School of Science, PO Box 11100, 00076 Aalto, Finland}
\author{Dani\"{e}l Vanmaekelbergh}
\affiliation{Condensed Matter and Interfaces, Debye Institute for Nanomaterials Science,
Utrecht University, PO Box 80000, 3508 TA Utrecht, the Netherlands}
\author{Peter Liljeroth}
\email[]{P.Liljeroth@uu.nl}
\affiliation{Condensed Matter and Interfaces, Debye Institute for Nanomaterials Science,
Utrecht University, PO Box 80000, 3508 TA Utrecht, the Netherlands}
\affiliation{Low Temperature Laboratory, Aalto University School of Science, PO Box 15100, 00076 Aalto, Finland}

\date{\today}

\begin{abstract}
Epitaxial graphene grown on transition metal surfaces typically exhibits a moir\'e pattern due to the lattice mismatch between graphene and the underlying metal surface. We use both scanning tunneling microscopy (STM) and atomic force microscopy (AFM) to probe the electronic and topographic contrast of the graphene moir\'e on the Ir(111) surface. STM topography is influenced by the local density of states close to the Fermi energy and the local tunneling barrier height. Based on our AFM experiments, we observe a moir\'e corrugation of 35$\pm$10 pm, where the graphene-Ir(111) distance is the smallest in the areas where the graphene honeycomb is atop the underlying iridium atoms and larger on the fcc or hcp threefold hollow sites.
\end{abstract}


\maketitle

Epitaxial graphene can be grown on many transition metal surfaces using chemical vapor deposition (CVD).\cite{Wintterlin2009,Li2009,Reina2009} This process gives ready access to high-quality, large scale, graphene monolayers on surfaces where graphene growth is self-terminating (e.g. Cu, Ir, Pt).\cite{Wintterlin2009,Li2009,Bae2010} These layers can be characterized by surface science techniques and if necessary, transferred onto other substrates for further processing. The different metal surfaces can be coarsely classified based on how strongly the graphene layer interacts with the underlying metal substrate.\cite{Preobrajenski2008} For example, Ir(111) and Pt(111) surfaces interact weakly with the graphene layer and consequently, graphene still exhibits linear Dirac-like dispersion characteristic of isolated graphene.\cite{Preobrajenski2008,Pletikosic2009,Sutter2009B} On the other hand, on Ru(0001) and Ni(111) surfaces, the graphene band structure is strongly modified.\cite{Preobrajenski2008,Sutter2009A} While the CVD growth occurs epitaxially, the lattice mismatch between graphene and the metal substrate gives rise to a moir\'e pattern that is observed on most metal surfaces (notably Ir(111)\cite{NDiaye2006,NDiaye2008,Loginova2009,Coraux2009}, Rh(111) \cite{Wang2010}, Ru(0001)\cite{Parga2008,Martoccia2008,Moritz2010} and Cu(111)\cite{Gao2010}).

This moir\'e pattern can be readily observed by scanning tunneling microscopy (STM).\cite{NDiaye2006,Parga2008,Gao2010} However, STM images do not directly probe the topography of the surface; instead, the STM tip traces constant integrated local density of states (LDOS) surfaces at energies close to the Fermi level.\cite{Giessibl2003,Hofer2003} This causes the contrast and apparent corrugation of the graphene moir\'e on Ir(111) to depend on the STM imaging conditions.\cite{NDiaye2006,NDiaye2008} It is not a priori clear which STM images correspond to the actual topography of the surface.

The use of a quartz tuning-fork force sensor in the QPlus configuration has made it possible to carry out non-contact atomic force microscopy (nc-AFM) in the frequency modulation mode with small tip oscillation amplitudes. This allows concurrent STM experiments, where the performance of the STM mode is not compromised by the tip oscillation or small force constant of the AFM cantilever.\cite{Giessibl2003,Hembacher2003,Hembacher2005,Ternes2008,Gross2009,Baykara2010} We have used this technique and performed both low-temperature AFM and STM measurements on epitaxial graphene monolayers on Ir(111) aimed at understanding the contributions of actual topography, charge transfer giving rise to local variations in the tunneling barrier height and contact potential difference, and variations of the LDOS on the observed moir\'e pattern. These techniques give independent information on the surface topography, which allows separating electronic and topographic effects.

\begin{figure*}
\includegraphics[width=.95\textwidth]{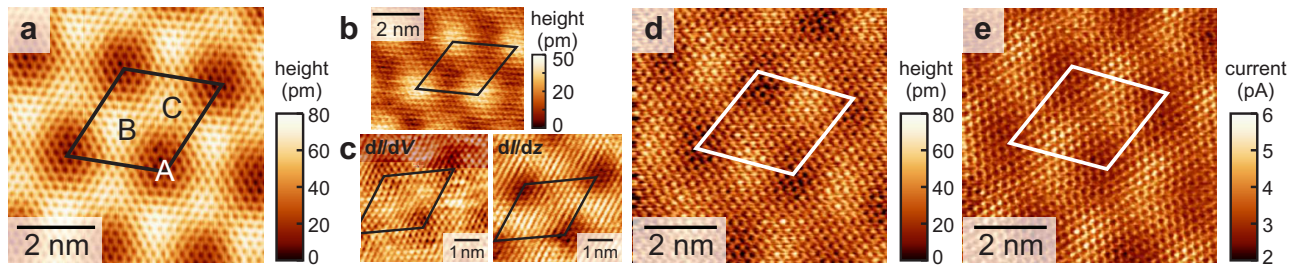}
\caption{(Color online) (a) Constant-current STM topography image of epitaxial graphene on Ir(111) acquired at $V_\text{bias}=0.3$ V and set-point current of 0.3 nA. The line indicates the moir\'e unit cell and the three inequivalent areas within it are denoted by A, B and C. (b) High-bias STM image taken at 0.45 V / 1 nA showing the inverted moir\'e pattern. (c) Constant-current d$I$/d$V_\text{bias}$ and d$I$/d$z$ maps recorded at a bias of 0.05 V. (d) Constant frequency shift nc-AFM image with $\Delta f=-45$ Hz and $V_\text{bias}=0.01$ V. (e) Average current over the tip oscillation cycle measured simultaneously under AFM feedback.}\label{AFMtopo}
\end{figure*}

The graphene was grown on Ir(111) by CVD from ethylene.\cite{Coraux2009} The Ir(111) surface was first cleaned by repeated cycles of 3 kV Ar$^+$-ion sputtering at room temperature followed by flashing to 1400 K and annealing at 1200 K. After the last annealing cycle, the sample was first annealed for 3 min at 800 K in $1 \cdot 10^{-7}$ mbar O$_2$, then flashed to 1400K before starting the CVD process. In order to ensure the formation of a full graphene monolayer, the sample was exposed to $5 \cdot 10^{-7}$ mbar ethylene at 1250 K for 100 s. The sample was then transferred ex situ to the low-temperature STM/AFM system (LT-STM equipped with QPlus force sensor, Omicron Nanotechnology Gmbh). Prior to the STM/AFM measurements, the sample was cleaned by heating to 800 K. All STM/AFM experiments were carried out in ultra-high vacuum (base pressure $<10^{-10}$ mbar) and at low-temperature ($T=4.7$ K). The QPlus sensor used for the frequency modulation nc-AFM experiments had a spring constant $k$ of 1800 N/m, resonance frequency $f_0$ of ca. 24 kHz and a quality factor of 18800. We used PtIr tips and the tip oscillation amplitude was set to 5 \AA. Bias voltage ($V_\text{bias}$) was applied on the sample with respect to the tip. d$I$/d$V_\text{bias}$ and d$I$/d$z$ signals were recorded with a lock-in amplifier by applying a small sinusoidal variation to the bias voltage or the $z$-piezo position, respectively.

Figure \ref{AFMtopo}a shows a constant-current STM topography image of a graphene monolayer on Ir(111). In addition to the atomically resolved hexagonal graphene structure, a moir\'e pattern with a period of 2.5 nm is clearly visible. It has been shown previously that this superstructure preferentially orients along the atomic rows of graphene, which is also the case in Figure \ref{AFMtopo}a.\cite{NDiaye2008} The apparent peak to peak (p-p) corrugation of the moir\'e pattern is 50 pm. The unit cell of the moir\'e is indicated by the solid line, and the three inequivalent areas (with respect to the registry with the Ir(111) lattice) are indicated by A, B, and C. It has been previously suggested that they correspond to areas where the graphene honeycomb is centered on the underlying Ir atoms (atop, A), or on the fcc (B) or hcp (C) threefold hollow sites.\cite{NDiaye2006,NDiaye2008} We observed that the apparent corrugation depends on tip conditions and systematically on the bias voltage, in agreement with earlier results.\cite{NDiaye2008} STM images at a higher bias ($\gtrsim 0.5$ V, Figure \ref{AFMtopo}b) exhibit inverted moir\'e contrast (region A becomes bright) compared to low-bias images.

In addition to standard STM imaging, we can get further information on the local electronic properties by mapping out the LDOS ($\propto$ d$I$/d$V_\text{bias}$) and the tunneling decay constant $\kappa$ ($\propto$ d$I$/d$z$) signals in the constant-current mode. These quantities vary over the moir\'e pattern as shown in Figure \ref{AFMtopo}c. Both LDOS and $\kappa$ are lower at region A of the moir\'e.

As STM imaging is clearly influenced by electronic effects, we carried out low-temperature nc-AFM experiments in the constant frequency shift mode to probe the surface topography of graphene monolayer on Ir(111). Figure \ref{AFMtopo}d shows a typical AFM topography image (the image was not recorded on the precise location of the image shown in Figure \ref{AFMtopo}a). In addition to the contrast on the atomic scale, we obtain a moir\'e pattern with similar contrast as in STM imaging and an apparent corrugation of ca. 30 pm. Careful inspection of Figure \ref{AFMtopo}d reveals variations in the atomic scale contrast. On the bright areas of the moir\'e pattern, the carbon atoms are imaged as depressions, in line with the earlier atomically resolved images of carbon nanotubes.\cite{Ashino2004} The observed contrast changes on the dark areas, indicating that the tip-graphene distance is different on the different regions of the moir\'e pattern (see below).

The simultaneously measured tunneling current during the nc-AFM imaging is shown in Figure \ref{AFMtopo}e. It again shows both atomic and moir\'e contrast, where the low current regions are aligned with the depressions in the topographic image. Note that there is a shift between the AFM image and simultaneously measured tunneling current. It is likely that the tip has an impurity (atom) that does not contribute to the current but has an effect on the measured frequency shift. Alternatively, an asymmetric tip apex can cause the shift between AFM and average current images.\cite{Sugimoto2010} We find the same qualitative moir\'e contrast with different tips and on different locations of the sample.

The tip-sample interaction causes a shift $\Delta f$ in the resonance frequency of the cantilever. At small tip oscillation amplitudes, the measured detuning is directly proportional to force gradient, $\Delta f= -f_0 /(2k) (\partial F_\text{ts}/\partial z)$, where $F_\text{ts}$ is the total interaction force between tip and sample.\cite{Giessibl2003} Different forces contribute to $F_\text{ts}$, the most relevant in our experiment being quantum mechanical forces between the tip apex and the surface (Pauli repulsion, chemical bonding), vdW interactions between the tip and graphene and the tip and the Ir substrate, and electrostatic forces.\cite{Giessibl2003,Hofer2003,Gross2009} AFM topography might also be affected by chemical inhomogeneity of the surface (different regions of the graphene moir\'e are known to have different chemical reactivities\cite{NDiaye2006,Wang2010,Parga2008}). However, we have observed the same qualitative moir\'e contrast with different tip terminations, consistent with the expected weak interaction between graphene and the Ir(111) surface.

\begin{figure}
\includegraphics[width=0.4\textwidth]{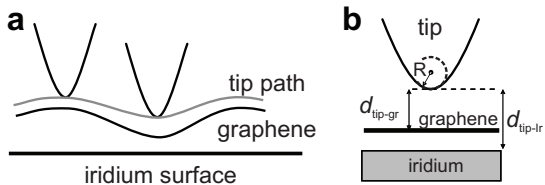}
\caption{(a) Schematic representation of the tip movement over graphene. (b) Variables used in Eq. \ref{df_tot}.}
\label{AFM_schem}
\end{figure}

Experimental results based on photoelectron spectroscopy (ARPES and XPS) and ab initio calculations show that the interaction between graphene and iridium is weak.\cite{NDiaye2006,Preobrajenski2008,Feibelman2008,Pletikosic2009} Theoretical calculations give an average graphene-Ir(111) distance of about 3.9 \AA\ (GGA-DFT) or 3.42 \AA\ (LDA-DFT).\cite{NDiaye2006,Feibelman2008} It is well-known that GGA underestimates and LDA overestimates binding in systems where vdW interactions are important. Recent DFT calculations using vdW-corrected functionals have found a binding distances of 3.6--3.7 \AA\ for graphene on weakly interacting metals (e.g. Pt).\cite{Vanin2010} Despite the large binding distance, it is important to realize that the vdW forces between the tip and the sample are sufficiently long range to include contributions from the iridium substrate. In the attractive regime, the background vdW from the Ir substrate results in increased attraction in the area A of the moir\'e, which causes the AFM feedback to increase tip-sample distance in order to keep $\Delta f$ constant. Hence, the AFM corrugation underestimates the real topographic corrugation of the graphene moir\'e as illustrated in Figure \ref{AFM_schem}a.

We will now model this effect within the small amplitude approximation (detuning proportional to force gradient). This approximation is valid if the force gradient is roughly constant throughout the
oscillation cycle of the tip. We use relatively small oscillation amplitudes and consider here only the vdW interactions that are relevant for the AFM observation of the moir\'e pattern on an otherwise chemically homogeneous surface. The chemical interactions between the tip apex and the surface only act at very short distances (much smaller than our tip oscillation amplitude) whereas the long-range electrostatic force remains almost constant over the oscillation cycle assuming that neither the tip nor the surface is charged. Consequently, these forces make only a minor contribution to the observed AFM response.

The total vdW force felt by the tip (modeled as a paraboloid $z=x^2/(2R)$, where $R$ is the tip radius) can be calculated by integrating the vdW potential $w_\text{vdw} = -4\epsilon (\sigma / r)^6$ over the tip and Ir bulk, and tip and two-dimensional graphene layer.\cite{Hofer2003} We assume that the vdW interaction between the tip and Ir substrate is not screened by the graphene layer. Hence, we obtain an estimate of the upper limit of the background vdW contribution. The derivative of the total force is then proportional to the detuning $\Delta f$ of the tip
\begin{equation}
\Delta f = -\frac{f_0R}{2k}\left ( \frac{\sqrt{A_\text{tip} A_\text{Ir}}}{3d^3_\text{tip-Ir}} + \frac{\sqrt{A_\text{tip}
A_\text{HOPG}}}{d^4_\text{tip-gr}l_\text{HOPG}} \right ) \label{df_tot}
\end{equation}
where $A_i = 4 \pi^2 \epsilon_i \rho_i^2 \sigma_i^6$ is the Hamaker constant and $d_\text{tip-Ir}$ and $d_\text{tip-gr}$ are the tip-Ir and tip-graphene distances corresponding to the midpoint of the tip oscillation cycle. In the case of a two-dimensional layer, the vdW force depends on the surface atom density rather than the volume density. We take this into account by using the HOPG Hamaker constant $A_\text{HOPG}$ and the layer density $l_\text{HOPG}$.

\begin{figure}
\includegraphics[width=0.33\textwidth]{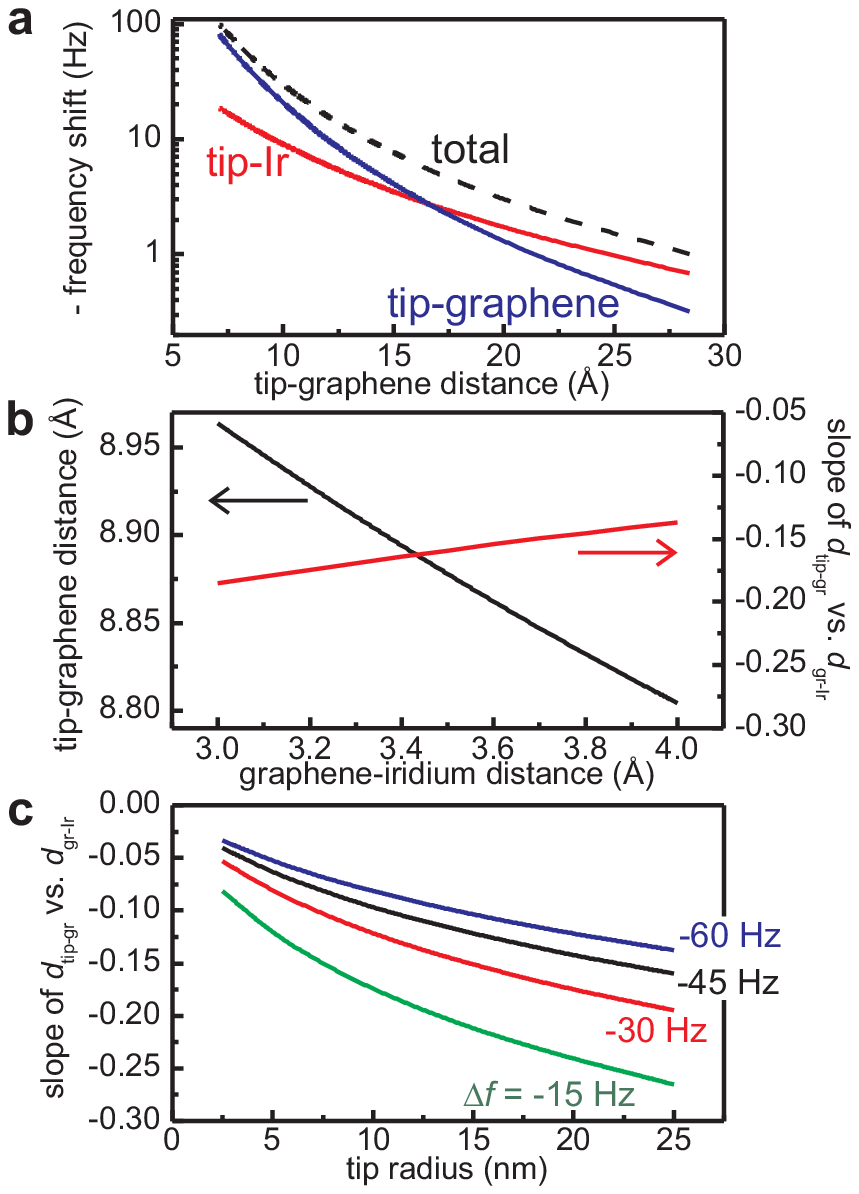}
\caption{(Color online) Theoretical modeling of the vdW forces in the AFM tip-graphene-Ir substrate system. (a) The contributions of graphene and iridium substrate to the total $\Delta f$. (b) Tip-graphene distance as a function of graphene-Ir distance calculated with $\Delta f = -45$ Hz. The corresponding slope is plotted on the right $y$-axis. (c) Slope $s$ of the $d_\text{tip-Ir}$ vs. $d_\text{tip-gr}$ curve for different tip radii and different $\Delta f$ (indicated in the figure). Panel (a) and (b) are calculated with $R=10$ nm. $A_\text{HOPG}=3.42\cdot 10^{-19}$ J, $A_\text{tip}=A_\text{Ir}= 4 \cdot 10^{-19}$ J \cite{visser1972,israel1992}, and $l_\text{HOPG} = 1/335.4$ pm were used in all calculations.}\label{AFM_theor}
\end{figure}

In Eq. \ref{df_tot} the first term is the detuning caused by the tip-Ir vdW force and the second term the tip-graphene vdW force. The relation of these terms is illustrated in Figure \ref{AFM_theor}a, where we plot them (and the total $\Delta f$) as a function of the tip-graphene distance. The contribution from the Ir substrate increases and even becomes the dominant term at large distances.

We have solved the tip-graphene distance $d_\text{tip-gr}$ from Eq. \ref{df_tot} numerically as a function of the graphene-Ir distance $(d_\text{tip-Ir}-d_\text{tip-gr})$ (Figure \ref{AFM_theor}b). $d_\text{tip-gr}$ changes almost linearly over a reasonable range of graphene-Ir distances. The slope $s$ of this curve represents the underestimation of corrugation on the moir\'e due to the background vdW forces from the Ir bulk. Thus the real geometric corrugation of the moir\'e is given by $a_\text{real} = (1 - s) a_\text{AFM}$. The Hamaker constants have a fairly small effect on the slope whereas $R$ and $\Delta f$ have quite a significant effect. This is illustrated in Figure \ref{AFM_theor}c, where we plot the average slope in the range of $d_\text{gr-Ir}$ between 3--4 \AA. This effect can be easily understood by the fact that $\Delta f$ and $R$ determine the absolute tip-graphene distance, which governs the proportion of the vdW force from the bulk Ir with respect to the total force. While our estimation of $d_\text{tip-gr}$ depends on $R$, which is difficult to estimate independently, we can also estimate $d_\text{tip-gr}$ based on the simultaneously recorded tunneling current. Taking into account the tip oscillation\cite{Giessibl2003} and using a measured value of the tunneling decay constant $\kappa\approx$ 0.55 {\AA}$^{-1}$, we can extrapolate the distance to point-contact. This procedure gives average tip-graphene distance of 7.2 {\AA} in Figure \ref{AFMtopo}d.

The data shown in Figure \ref{AFM_theor}c shows that even though the background vdW from the Ir bulk affects the apparent AFM corrugation, the effect is rather small. The correction factor is 10-20\% for reasonable tip radii and at sufficiently negative $\Delta f$, which yields 35$\pm$10 pm as our estimation for the actual moir\'e corrugation. It should be noted that this model does not take the actual shape of the corrugation of the moir\'e into account. It only corrects the measured corrugation by the background vdW from the Ir bulk. Due to the relatively equal scale of curvature of the tip and the moir\'e, the corrugation may be slightly further underestimated by AFM measurements.

The increase in the tip-graphene distance caused by the increased background vdW interaction between the tip and the Ir substrate in the area A of the moir\'e reduces the tunneling current as seen in Figure \ref{AFMtopo}c. However, if we use the estimated changes in $d_\text{tip-gr}$, we obtain a current variation of only ca. 10 \%. On the other hand, the experimentally measured variation is much larger, ca. a factor of 2. Apart from the $d_\text{tip-gr}$, the tunneling current is influenced by the LDOS and the decay constant $\kappa$. Both of these quantities vary over the moir\'e pattern as shown in Figure \ref{AFMtopo}c. These quantities have opposite effects on the tunneling current: the larger the LDOS, the larger the current. On the other hand, the larger the decay constant, the smaller the current as it is proportional to $\exp(-2\kappa d_\text{tip-gr})$. Our observations then imply that at small bias, the reduced LDOS in the region A of the moir\'e is (mostly) responsible for the reduced tunneling current in the simultaneously measured tunneling current images under AFM feedback. This conclusion naturally does not hold for increased bias ($\gtrsim 0.5$ V) where the STM contrast of the moir\'e pattern is inverted.

We can relate the STM to AFM results by switching in situ back and forth between STM and AFM feedback. Qualitatively, the moir\'e contrast is the same between AFM and low bias STM images (dark depressions in a bright background). We do not have a direct measure of the registry between the moir\'e unit cell and the underlying iridium lattice. However, comparison of our STM results with the STM-based graphene adsorption site determination\cite{NDiaye2006,Feibelman2008,NDiaye2008} relates the regions of the moir\'e unit cell to areas where the graphene honeycomb is centered atop the underlying Ir atoms (A), or on the fcc (B) or hcp (C) threefold hollow sites. Thus, our STM and AFM measurements seem to imply that graphene-Ir(111) distance is the smallest on atop sites (region A), and larger on fcc and hcp sites (regions B and C). This is surprising and in contrast to other graphene-metal systems.

In conclusion, we have carried out simultaneous low temperature AFM and STM experiments on an epitaxial graphene monolayer on the Ir(111) surface. These experiments shed light to the structure of the graphene moir\'e on the Ir(111) surface. While STM experiments are dominated by electronic effects, nc-AFM provides a qualitatively correct image of the surface topography. A more quantitative estimation of the moir\'e corrugation based on the AFM experiments would require accounting for the background vdW interaction between the tip and the metallic substrate. Although in the present case of graphene on Ir(111) the background effect is small, it has to be considered in principle for quantitative topography of atomically thin two-dimensional layers deposited on solid substrates.

\begin{acknowledgments}
This research was supported by the Academy of Finland (Projects 117178 and 136917) and NWO (Chemical Sciences, Vidi-grant 700.56.423).

\end{acknowledgments}

\end{document}